\documentclass[aps,preprint,showkeys]{revtex4}
\usepackage{amsfonts}
\usepackage{graphicx}
\usepackage{amssymb}
\begin{document}
\title
{ Ferroelectric Mott-Hubbard phase in organic conductors.}
\author {S. Brazovskii}
\affiliation{LPTMS-CNRS, Orsay, France {\rm and} Landau Institute, Moscow, Russia}
\author{P. Monceau}
\affiliation{CRTBT-CNRS, Grenoble {\rm and}  LLB CEA-CNRS, Saclay, France}
\author{ F. Nad}
\affiliation{Institute of Radioelectronics, Russian Ac. Sci., Moscow, Russia}
\bigskip
 \bigskip
\begin{abstract}
We present key issues of  related  phenomenons of the Ferroelectricity and the
Charge Disproportionation  in organic metals. In  $(TMTTF)_{2}X$ the dielectric
susceptibility $\varepsilon$ demonstrates clear cases of the ferroelectric and
anti-ferroelectric phase transitions.  Both $\varepsilon$ and $\sigma$ prove
independence and occasional coexistence of \lq{structurless}\rq\ ferroelectric
transitions and usual \lq{anionic}\rq\, ones.  Their sequence gives access to
physics of three types of solitons emerging upon cooling via several steps of
symmetry breaking. The theory invokes a concept of the Combined Mott-Hubbard State
which focuses upon weak processes of electronic Umklapp scattering coming from both
the build-in nonequivalence of bonds and the spontaneous one of sites. We  propose
that the charge ordering in its form of the  ferroelectricity exists hiddenly even
in the $Se$ subfamily $(TMTSF)_{2}X$, giving rise to the unexplained yet low
frequency optical peak and the enhanced pseudogap.
\end{abstract}
\bigskip
\published[Proceedings of ICSM 2002 \cite{ICSM-02}.] {\hskip 5mm
Written in October 2002 for \cite{BMN-02}}
\bigskip

\maketitle

\section{Ferroelectricity in organic conductors.}
\begin{figure}[tbp]
\begin{center}
\includegraphics*[width=7.5cm]{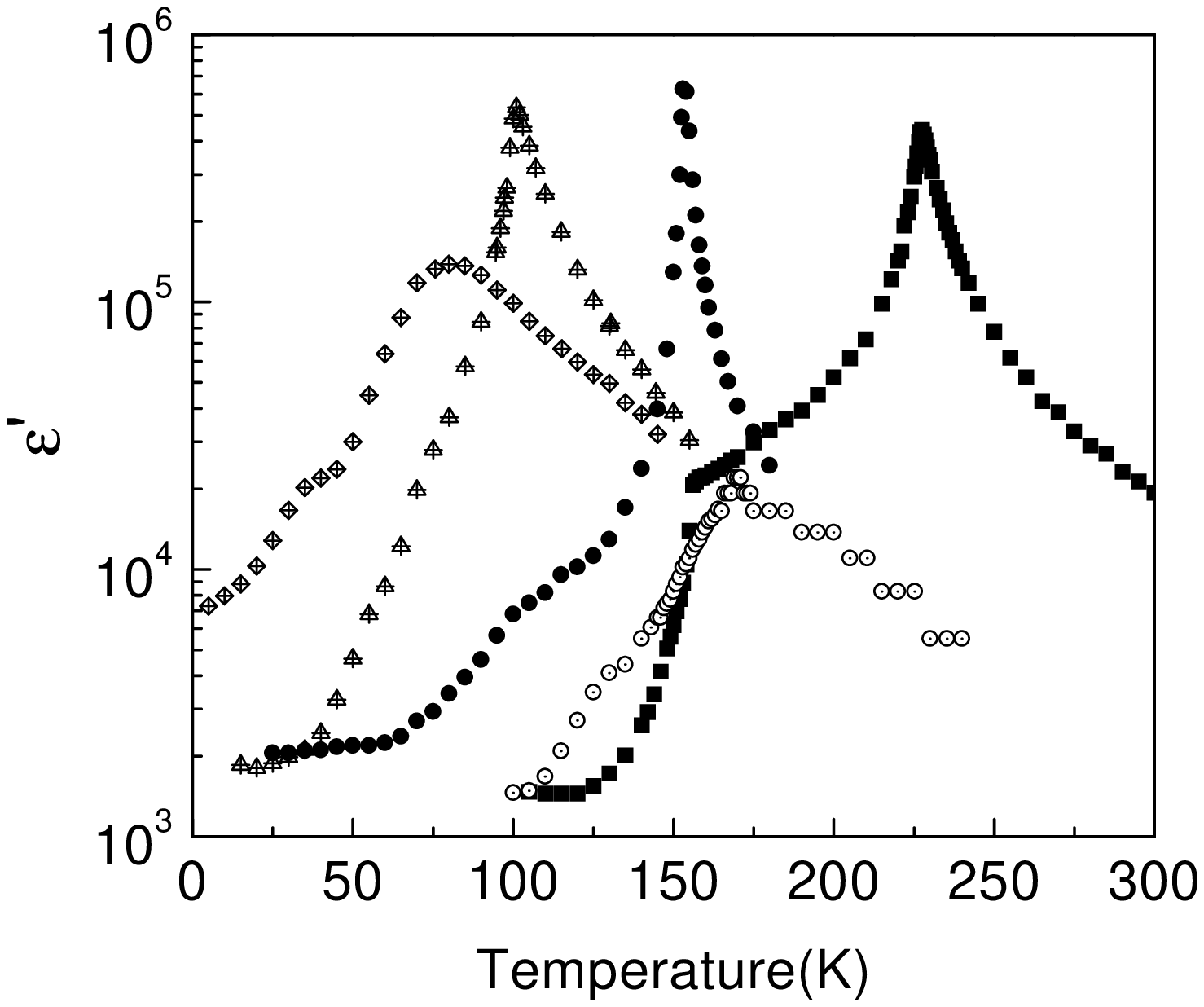}
\end{center}
\caption{$\log\varepsilon$ vs $T$ for FE cases $X=PF_{6}$
($\Diamond$), $AsF_{6}$ ($\triangle $), $SbF_{6}$ ($\bullet$),
$ReO_{4}$ ($\blacksquare $) and for the anti-FE $X=SCN$ ($\odot$).}
\label{epsilon}
\end{figure}

The family of quasi one-dimensional organic superconductors (the
Bechgard - Fabre salts $(TMTSF)_{2}X, (TMTTF)_{2}X $)
demonstrates, at low temperatures $T$, transitions to almost all
known electronic phases, see \cite{jerome}. At higher $T=T_{ao}$,
usually there is also a set of weak structural transitions of the
\lq{anion orderings}\rq\, (AOs) which are slight arrangements of
chains of counterions $X$ \cite{anions}. At even higher $T\approx
T_{0}$, also other \lq{structurless}\rq\,   transitions
\cite{lawersanne} were observed sometimes in the $TMTTF$ subfamily
but they were not explained and left unattended.
 Recently their mysterious nature has been elucidated by discoveries of the huge
 anomaly in the dielectric susceptibility
$\varepsilon$ \cite{prl,nad} (see Fig. \ref{epsilon}) and of the
charge disproportionation (CD) seen by the NMR  \cite{brown}. The
new displacive instability and the usual \lq{orientational}\rq\,
AOs seem to be independent, as proved by finding their sequence in
$(TMTTF)_{2}ReO_{4}$ \cite{nad}.

\begin{figure}[tbp]
\begin{center}
\includegraphics*[width=7.5cm]{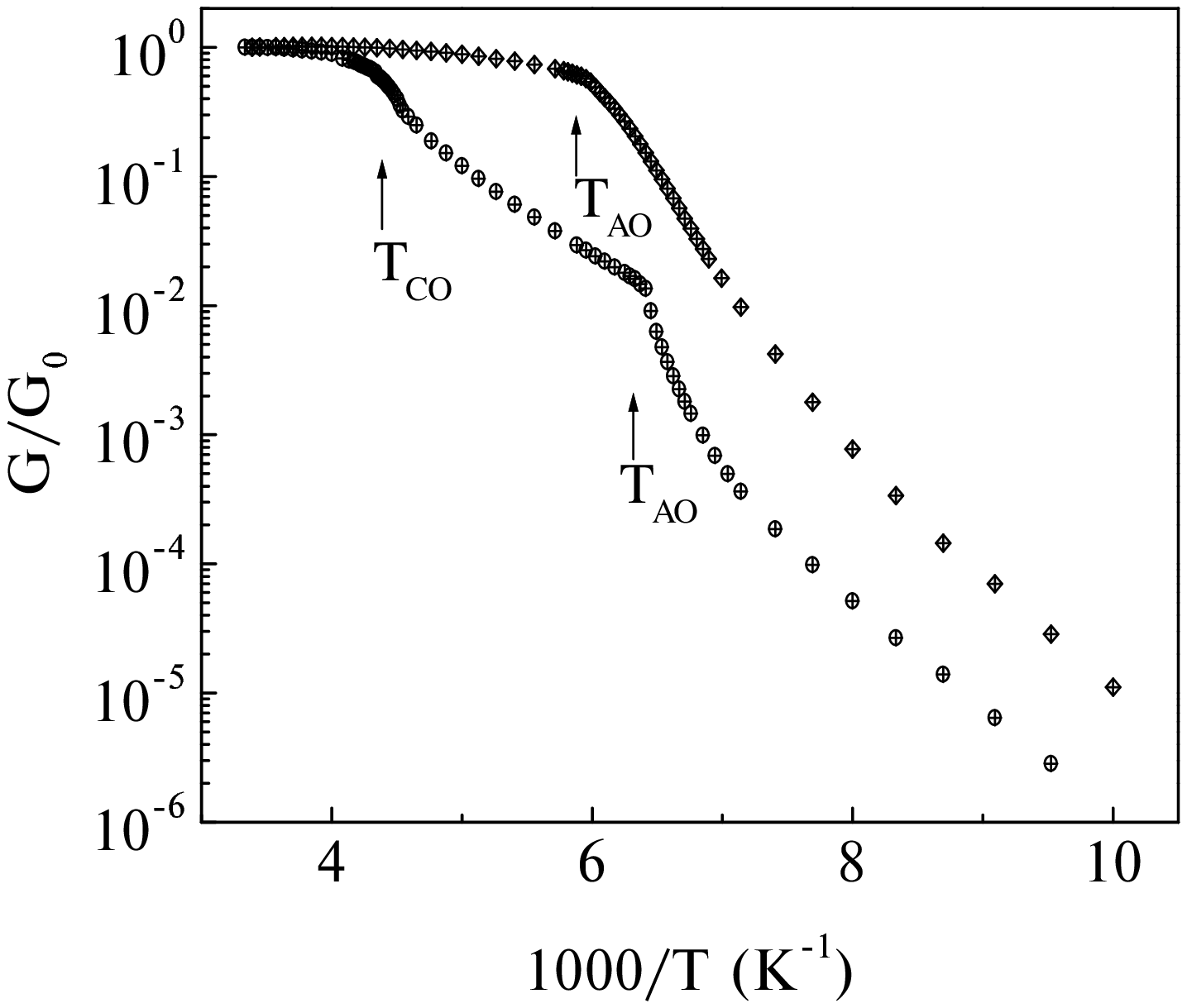}
\end{center}
\caption{Ahrenius plots for the normalized conductivity $G/G_{RT}$ (at $1kHz$)
for $X=ReO_{4}$ ($\circ $)  and $X=SCN$ ($\Diamond$).}
\label{cond}
\end{figure}

The phase transition was interpreted \cite{prl} as the least expected one: to
the Ferroelectric (FE) state, which was proved by the clear-cut fitting of the
anomaly in $\varepsilon (T)$ to the Curie law (see the figure in \cite{prl}).
The FE transition is followed by a fast formation, or the steep increase, of
the conductivity gap $\Delta$ (see the Fig.\ref{cond}) with no signs of a spin
gap formation. Hence we deal with a surprising FE version of  the  Mott-Hubbard
state which usually was associated rather with magnetic  orderings. The FE
transition in $(TMTTF)_{2}X$ is a  very particular, bright manifestation of a
more general phenomenon of the CD, which already has been predicted in
\cite{seo-97} and now becomes recognized as a common feature of organic and
some other conductors \cite{fukuyama}.

 The anomalous diverging polarizability is coming from the electronic system,
even if ions are very important to choose and stabilize the long
range $3D$ order. Thus the theory \cite{prl} suggests that the
collective singular contribution $\varepsilon\sim
({\omega_{p}}/{\Delta )^{2}}/|T/T_{0}-1|$ ($\omega _{p}$ is the
metallic plasma frequency) develops upon the already big intergap
contribution $\sim ({\omega _{p}}/{\Delta )^{2}\sim 10}^{3}$ which
is also seen as the background at the Fig.\ref{epsilon}. Typical
plots at the Fig.\ref{epsilon} demonstrate very sharp (even in the
$\log$ scale !) FE peaks, while the  anti-FE case of $X=SCN$ shows
a  smoother maximum. At the subsequent Ist order AO transition in
the case of $X=ReO_{4}$, $\varepsilon$ drops down which might be
caused by  the increase of $\Delta$ seen at the conductivity plot
at the Fig.\ref{cond}. Finally the IInd order spin- Peierls
transition in the case of $X=PF_{6}$ shows up only as a shoulder.
All that seems to collaborate towards a consistent picture.

Already within the nonperturbed crystal structure at $T>T_{0}$, the tiny dimerization
of bonds by anions $X$ provokes the dielectrization \cite{SB-VY}, see more references
in \cite{brazov-01}. The counting of the mean electronic occupation changes from
$1/2$ per molecule to $1$ per their dimer which opens (according to Dzyaloshinskii \&
Larkin, Luther \& Emery) the route to the Mott-Hubbard insulator. At $T<T_{0}$ the CD
adds more to the charge gap $\Delta$ which is formed now by \emph{joint effects of
alternations of bonds and sites}. The conductivity $G$ plots at the Fig.\ref{cond}
show this change by kinks at $T=T_{0}$ turning down to much higher activation
energies at low $T$. The steepness of $G$ just below $T_{0}$ reflects the growth of
the CD contribution to the gap $\sqrt{T_0-T}$, see below.

None of these two perturbations changes the unit cell of the zigzag stack which
basically contains two molecules, hence $q_{\parallel}=0$ (${\bf
q}=(q_{\parallel},q_{\perp})$ is the CD wave vector). Their sequence lifts the mirror
and then the inversion symmetries which must lead to the on-stack electric
polarization. By a good fortune, the 3D pattern of the CD appears in two, anti-FE and
FE, forms: i) antiphase between stacks, $q_{\perp}\neq 0$ which allows for its
structural identification \cite{anions} (found only for $X=SCN$); and ii) inphase
(${\bf q}=0$ hence the \lq{structurless}\rq\, character) which is the macroscopic FE
state typically observed today \cite{nad}. Both types are the same paramagnetic
insulators (the MI phase of \cite{SB-VY}) as we can see in Fig.\ref{cond} and in
other examples \cite{nad}; also their CD shows up similarly in the NMR splitting
\cite{brown}.

\section{Symmetry breakings and emergence of solitons.}
The microscopic theory for  the "combined Mott-Hubbard state"
\cite{prl} accounts for two \lq{orthogonal}\rq\, contributions,
$U_{b}$ and $U_{s}$, to the Umklapp scattering of electrons which
come from two symmetry breaking effects: the built-in
nonequivalence of bonds and the spontaneous one of sites
\cite{matv}. The appearance of $U_{s}$ is regulated by a single
parameter $\gamma$ ($\gamma _{\rho }$ of \cite{SB-VY}, $K_{\rho }$
of today's convention) which collects all information about
electronic interactions. The spontaneous CD $U_{s}\neq 0$ requires
that $\gamma <1/2$, far enough from $\gamma =1$ for noninteracting
electrons. The magnitude $|U_{s}|$ is determined by a competition
between the electronic gain of energy and its loss $\sim
U_{s}^{2}$ from the lattice deformation and charge redistribution.
The $3D$ ordering of signs $U_{s}=\pm |U_{s}|$ discriminates
between the FE and anti-FE states.

While the earliest theoretical approach \cite{seo-97} applies well
to a generic CD, here the pronouncedly $1D$ electronic regime
calls for a special treatment \cite{prl} which also needs to be
well suited for the FE transition. It is done in terms of
electronic phases $\varphi$ and $\theta$ defined as for the CDW
order parameters $\sim \exp(i\varphi)\cos\theta$) such that
$\varphi ^{\prime }/\pi $ and $\theta ^{\prime }/\pi $ count local
concentrations of the charge and the spin. Beyond the energies of
charge  and spin polarizations $\sim(\hbar
v_{F}/\gamma)(\varphi^{\prime})^{2}$ and $\sim\hbar
v_{F}(\theta^{\prime})^{2}$, there is the commensurability energy
\[
H_{U}=-U_{s}\cos 2\varphi -U_{b}\sin 2\varphi =-U\cos (2\varphi -2\alpha )
\]
where $U=\sqrt{U_{s}^{2}+U_{b}^{2}}$ and $\tan 2\alpha =U_{b}/U_{s}$. The gap
$\Delta$ is related to the total Umklapp amplitude $U$ via its renormalized value
$U\Rightarrow U^{\ast }\sim \Delta ^{2}/\hbar v_{F}$.

For a given $U_{s}$, the ground state is doubly degenerate between
$\varphi=\alpha$ and $\varphi =\alpha +\pi $ which allows for
phase $\pm\pi$  solitons with the energy $\Delta$ which are the
charge $\pm e$  spinless particles, the (anti)holons observed in
conductivity  at both $T\gtrless T_{0}$. Also $U_{s}$ itself can
change the  sign between different domains of ionic displacements.
Then the electronic  system must also adjust its ground state from
$\alpha$ to $-\alpha $ or to  $\pi -\alpha$. Hence the FE domain
boundary $U_{s}\Leftrightarrow -U_{s}$  requires for the phase
$\alpha$-solitons of the increment $\delta \varphi=-2\alpha $ or
$\pi -2\alpha$ which will concentrate the non integer charge
$q=-2\alpha /\pi$ or $1-2\alpha /\pi $ per chain. Below $T_{0}$,
the $\alpha -$ solitons must be aggregated into domain walls
\cite{teber} which motion might be responsible for the observed
frequency dispersion of $\varepsilon$, and indeed it is more
pronounced bellow $T_0$, see \cite{nad}. But well above $T_{0}$
they can be seen as individual particles, charge carriers. Such a
possibility requires for the fluctuational $1D$ regime of growing
CD. It seems to be possible sometimes  in view of a strong
increase of $\varepsilon$ at $T>T_0$ even for the anti-FE case of
the $X=SCN$ which signifies the growing single chain
polarizability before the 3D interactions come to the game.

The subsequent AO of the tetramerization in $(TMTTF)_{2}ReO_{4}$ exhorts upon
electrons a CDW type effect thus adding the energy
$\sim\Delta_{\sigma}^{2}\cos(\varphi -\beta)\cos \theta$ (the  shift $\beta$, the
mixture of bond- and site CDWs, reflects the lack of the inversion symmetry below
$T_0$). It lifts the continuous $\theta$ invariance thus opening the spin gap
$\Delta_{\sigma}$ growing below $T_{ao}$. Moreover it lifts even the discrete
$\varphi\rightarrow\varphi +\pi$ invariance of $H_{U}$ thus prohibiting the $\pi$
solitons. But the joint invariance ($\varphi\rightarrow\varphi +\pi
,\theta\rightarrow \theta +\pi$) is still present giving rise to \emph{combined
topological solitons}. They are composed by the charge $e$ core  (with $\delta
\varphi =\pi$ within the length $\sim \hbar v_{F}/\Delta $) which is supplemented by
longer spin $1/2$ tails of the $\theta -$ soliton ($\delta \theta =\pi $ within the
length $\sim\hbar v_{F}/\Delta _{\sigma}$). These are the particles seen at the
conductivity plot of the Fig.\ref{cond} below $T_{ao}$.

\section{Comparison, perspectives.}
By now the revaluation concerns definitely only the $TMTTF$ subfamily. The $TMTSF$
compounds are highly conductive which today does not allow for difficult  experiments
either with the low frequency $\varepsilon$ or with the small NMR splitting.
Nevertheless the transition may be their, just being hidden or existing in a
fluctuational regime like for stripes in High-$T_{c}$ cuprates. When it is confirmed,
then the whole picture of intriguing abnormal metallic state \cite{jerome} will be
revised. The signature of the FE CD state may have been already seen in optical
experiments \cite{optics}. Indeed the Drude like peak appearing within the pseudogap
can be interpreted now as the optically active mode of the FE polarization; the joint
lattice mass will naturally explain its surprisingly low weight. Vice versa, the FE
mode must exist in $TMTTF$ compounds, which identification is the ultimate goal.
(Being overdamped near $T_0$, this mode must grow in frequency following the order
parameter as $\sim\varepsilon^{-1/2}$ which can yield about two orders of magnitude
at low $T$.) Even the optical pseudogap itself \cite{optics}, being unexpectedly big
for $TMTSF$ compounds with their less pronounced dimerization of bonds, can be
largely due to the hidden spontaneous dimerization of sites.

A popular interpretation \cite{optics} for optics of $TMTSF$ compounds neglects
the dimerization and relies upon the more generic 4- fold commensurability
effects originating higher order (8 particles) Umklapp processes. They give rise
to the energy  $\sim U_{4}\cos 4\varphi$ which stabilization would require for
ultra  strong $e-e$ repulsion corresponding to $\gamma <1/8$ in compare to our
moderate constraint $\gamma <1/2$. While not excluded in principle, this
mechanism does not work in $TMTTF$. We saw that even small increments of the
dimerization just below $T_{0}$ immediately transfer to the activation energy.

Finally recall also numerical studies which have been performed in response to
new discoveries. Usually they pass the test for the CD but they may fail in
finding the FE; e.g. the nonpolar(1100 type ordering has been claimed \cite{mazilo},
rather than the necessary 1010 one.

In conclusion, new events call for a revision of the existing  picture and
suggest new experimental and theoretical goals. The world of organic metals
becomes polarized and  disproportionated.

\textbf{Acknowledgements.} This work was partly supported by the RFFI grant
N 02-02-17263, the INTAS grant 2212, and by the twinning CRTBT --IRE grant
N 98-02-22061. S.B. acknowledges the hospitality of the ISSP, Tokyo
University.


\newpage
\begin{thebibliography}{99}
\bibitem{ICSM-02} Proceedings of the ICSM 2002, Synthetic Metals,
v. {\bf 137}/1-3 2003.

\bibitem{BMN-02} S. Brazovskii, P. Monceau and F. Nad in \cite{ICSM-02}, p. 1331.

\bibitem{jerome} C. Bourbonnais and D. Jerome
\textit{in Adv. Synth. Met.}, (Elsevier 1999) \textit{and}
\textbf{cond-mat}/9903101.

\bibitem{anions} J.-P. Pouget and S. Ravy,
J. Physique I, \textbf{6}  1505 (1996).

\bibitem{lawersanne}
R. Lawersanne, C. Coulon, B. Gallois, J.-P. Pouget and R. Moret,
J. Physique. Lett. {\bf 45}, L393 (1984); H.H.S. Javadi, R.
Lawersanne and A.J. Epstein, Phys. Rev. B{\bf 37}, 4280 (1988);
 C. Coulon, S.S.P. Parkinand R. Lawersanne , Phys. Rev. B{\bf 31}, 3583 (1985).

\bibitem{prl} P. Monceau, F. Nad, S. Brazovskii,
Phys. Rev. Lett., \textbf{86 } 4080 (2001) \textit{and}
\textbf{cond-mat}/0012237.

\bibitem{yamada} Proceedings of the ISCOM IV,
Synthetic Metals, v. {\bf 133-134} (2003).

\bibitem{nad} F. Ya. Nad, P. Monceau, C. Carcel and J. M. Fabre,
\textit{in} \cite{yamada}, p. 265,  \textit{and refs. therein}.

\bibitem{brown} D. S. Chow \textit{et al},
Phys. Rev. Lett. \textbf{85} 1698 (2000);
\textit{also }S. Brown \textit{in} \cite{yamada} \textit{and in} \cite{ICSM-02}.

\bibitem{seo-97} H. Seo and H. Fukuyama,
    J. Phys. Soc. Japan \textbf{66} 1249 (1997).

\bibitem{fukuyama} H. Fukuyama,
    \textit{in} \cite{yamada} \textit{and in} \cite{ICSM-02}.

\bibitem{SB-VY} S. Brazovskii and V. Yakovenko, Sov. Phys.: JETP
\textbf{62}  1340 (1985){\it and references therein}.

\bibitem{brazov-01} S. Brazovskii, \textit{in} \cite{yamada}, p.
301, \textit{and references therein}, \textit{see also}
\textbf{cond-mat}/0304076.

\bibitem{0.5}  Both contributions can be of the build in type as in the mixture
    $(TMTSF)_{0.5}(TMTTF)_{0.5}$, \textit{see} \cite{anions}.

\bibitem{matv}
Recall  the \lq{combined Peierls state}\rq\,
 in the $(AB)_{x}$ conjugated polymers:
S. Brazovskii, N. Kirova N. and S. Matveenko,
Sov. Phys. JETP \textbf{59}  434 (1984).

\bibitem{teber} S. Teber, J. Physics C, {\bf 14}, 7811 (2002).

\bibitem{optics}
A. Schwartz, \textit{et~al.}, Phys. Rev. B 58, 1261 (1998).

\bibitem{mazilo} S. Mazumdar, \textit{in} \cite{ICSM-02}.
\end{thebibliography}
\end{document}